\documentclass[aps,prd,preprint,eqsecnum,floats,epsf,superscriptaddress,nofootinbib]{revtex4}
\usepackage{epsfig}
\usepackage{graphicx}
\usepackage{extarrows}
\usepackage{subfigure}

\def\ov{\overline}
\def\be{\begin{eqnarray}}
\def\en{\end{eqnarray}}
\def\non{\nonumber}
\def\vp{\varepsilon}

\def\la{\langle}
\def\ra{\rangle}
\def\A{{\cal A}}
\def\B{{\cal B}}

\def\lsim{ {\ \lower-1.2pt\vbox{\hbox{\rlap{$<$}\lower5pt\vbox{\hbox{$\sim$}
}}}\ } }
\def\gsim{ {\ \lower-1.2pt\vbox{\hbox{\rlap{$>$}\lower5pt\vbox{\hbox{$\sim$}
}}}\ } }

\begin{document}

\font\el=cmbx10 scaled \magstep2{\obeylines\hfill March, 2022}
\vskip 1.0 cm

\title{Two- and three-body hadronic decays of charmed mesons involving a tensor meson}

\author{Hai-Yang Cheng}
\affiliation{Institute of Physics, Academia Sinica, Taipei, Taiwan 11529, ROC}

\author{Cheng-Wei Chiang}
\affiliation{Department of Physics, National Taiwan University, Taipei, Taiwan 10617, ROC}
\affiliation{Physics Division, National Center for Theoretical Sciences, Taipei, Taiwan 10617, ROC}

\author{Zhi-Qing Zhang}
\affiliation{Department of Physics, Henan University of Technology, Zhengzhou, Henan 450052, P.R. China}


\begin{abstract}
\small
\vskip 0.5cm
We study the quasi-two-body $D\to TP$ decays and the three-body $D$ decays proceeding through intermediate tensor resonances, where $T$ and $P$ denote tensor and pseudoscalar mesons, respectively.  We employ $D\to T$ transition form factors based upon light-cone sum rules and the covariant light-front quark model to evaluate the decay rates, with the former giving a better agreement with current data.  Though the tree amplitudes with the emitted meson being a tensor meson vanish under factorization approximation, contributions proportional to the tensor decay constant $f_T$ can be produced from vertex and hard spectator-scattering corrections.  We also investigate the finite-width effects of the tensor mesons and find that, contrary to three-body $B$ decays, the tensor-mediated $D$ decays are more seriously affected and the narrow width approximation has to be corrected.  More experimental data are required in order to extract information topological amplitudes associated with quasi-two-body $D\to TP$ decays.  Among the data, the $D^+\to f_2\pi^+$ and $D^+\to \ov K_2^{*0}\pi^+$ branching fractions are not self consistent and further clarification is called for.

\end{abstract}

\maketitle
\small

\pagebreak

\section{Introduction}

In this paper, we set to study the quasi-two-body $D\to TP$ decays and the three-body $D$ decays proceeding through intermediate tensor resonances, where $T$ and $P$ denote tensor and pseudoscalar mesons, respectively. The $D \to TP$ decays have been studied previously in Refs.~\cite{Katoch:1994zk,Munoz:1998sn,ChengTP,Cheng:SAT,Momeni:2019eow}.  In Ref.~\cite{Cheng:SAT}, we pointed out that the $D \to TP$ measurements poise a big problem for theory.  It appeared that the predicted branching fractions based on the factorization approach were at least two orders of magnitude smaller than data, even when the decays were free of weak annihilation contributions.  Calculations in Refs.~\cite{ChengTP,Cheng:SAT} were based on the the Isgur-Scora-Grinstein-Wise (ISGW) model~\cite{ISGW} (or its improved version ISGW2 model~\cite{ISGW2}) and the covariant light-front quark model (CLFQM)~\cite{CCH} for $D\to T$ transition form factors. Recently, these form factors have been evaluated using light-cone sum rules (LCSR) in Ref.~\cite{Momeni:2019eow}. It turns out that form factors obtained from LCSR are much larger than those found in the ISGW model or CLFQM.  Consequently, the discrepancy between theory and experiment gets improved.

As discussed in Refs.~\cite{ChengTP,Cheng:SAT}, one generally has two sets of distinct diagrams for each topology in $D\to TP$ decays.  For example, there are two external $W$-emission and internal $W$-emission diagrams, depending on whether the emitted particle is an even-party meson or an odd-parity one.  Following the convention in Refs.~\cite{ChengTP,Cheng:SAT}, we shall denote the primed amplitudes $T'$ and $C'$ for the case when the emitted meson is a tensor meson.  Since the tensor meson cannot be produced from the $V-A$ current, its vector decay constant vanishes identically.  Hence, we have set $T'=C'=0$ before in the na{\"i}ve factorization approach.  Nevertheless, as stressed in Ref.~\cite{Cheng:TP}, beyond the factorization approximation, contributions proportional to the decay constant $f_T$ defined in Eq.~(\ref{eq:fT}) below can be produced through vertex and spectator-scattering corrections in the QCD factorization (QCDF) approach~\cite{BBNS} for hadronic $B$ decays.  Hence, in this work we will generalize QCDF to charmed mesons to estimate the nonfactorizable effects in $D\to TP$ decays.

There are four $D\to T$ transition form factors induced from the $(V-A)$ current, $A_0, A_1,A_2$ and $V$ parametrized  in Refs.~\cite{Cheng:TP,Wang:2010ni}, or $k, b_+, b_-$ and $h$ defined in the ISGW model (see Eq.~(\ref{eq:FFinISGW}) below).  The latter four form factors were calculated in CLFQM with the results listed in Table~VI of Ref.~\cite{Cheng:SAT}.
However, as pointed out in Ref.~\cite{CCH}, the form factor $k(q^2)$ at zero recoil was problematic as it did not respect heavy quark symmetry in the heavy quark limit.  This was the main reason why the calculated branching fractions of $D\to TP$ decays were at least two orders of magnitude smaller than data.  It was advocated in Ref.~\cite{CCH} that one might apply heavy quark symmetry to obtain the form factor $k(q^2)$.  In this work we will apply heavy quark symmetry to $D\to T$ transitions to see any improvement on $k^{DT}(q^2)$.

Very recently, the form factors of $P\to T$ transition were analyzed in Ref.~\cite{Chen:2021ywv} within the covariant light-front quark  model, which we will call CLFQM$_b$.  This time, the four form factors $A_0, A_1,A_2$ and $V$ were directly evaluated in CLFQM$_b$ in which some issues with the previous study of CLFQM were overcome.  We will consider the form factors obtained in this model as a benchmark for comparison.

This paper is organized as follows.  In Section~\ref{sec:status}, we review the current experimental status of the measurements of 3-body charmed meson decays that are relevant to our analysis.  We provide the information of flavor SU(3) classification, decay constants, and form factors for the $T$ mesons in Section~\ref{sec:properties}.  Section~\ref{sec:flavorapp} presents the so-called quark-diagram approach to the decays.  Each decay mode is decomposed in terms of quark diagrams characterized by their flavor topologies.  The goal is to see if current experimental data can be used to infer the magnitude and strong phase associated with each of the amplitudes.  In Section~\ref{sec:facDTP}, we study the flavor operators $a_{1,2}(M_1M_2)$ for $M_1M_2=TP$ and $PT$ within the framework of QCDF.  Under the factorization assumption, we compute the rate of each decay mode.  We also examine the finite width effects for certain decay modes in Section~\ref{sec:finitewidth}.   A summary of our findings is given in Section~\ref{sec:conclusions}.

\section{Experimental status \label{sec:status}}

It is known that three- and four-body decays of heavy mesons provide a rich laboratory for studying the intermediate state resonances.  The Dalitz plot analysis of three-body or four-body decays of charmed mesons is a very useful technique for this purpose.  We are interested in $D\to TP$  decays followed by $T\to P_1P_2$.  The results of various experiments are summarized in Table~\ref{tab:TPData}.  To extract the branching fraction for $D\to TP$, we apply the narrow width approximation (NWA)
 \be \label{eq:fact}
 \Gamma(D\to TP\to P_1P_2P)=\Gamma(D\to TP)_{\rm NWA}\B(T\to P_1P_2) ~.
 \en
Since this relation holds only in the $\Gamma_T\to 0$ limit, we put the subscript NWA to emphasize that $\B(D\to TP)$ thus obtained is under this limit.  Finite width effects in certain decays will be discussed in Section~\ref{sec:finitewidth}.  To extract the branching fractions of two-body decays of tensor mesons, we shall use~\cite{PDG}
\be
 \B(f_2(1270)\to\pi\pi) = (84.2^{+2.9}_{-0.9})\% ~, &&
 \B(f_2(1270)\to K\ov K) =(4.6^{+0.5}_{-0.4})\% ~,
 \non  \\
 \B(a_2(1320)\to K\ov K)=(4.9\pm 0.8)\% ~, &&
 \B(K_2^*(1430)\to K\pi)=(49.9\pm 1.2)\% ~. \non
 \en
The extracted branching fractions are shown in Table~\ref{tab:TPData} under the column $\B(D\to TP)_{\rm NWA}$.

Comparing Table~\ref{tab:TPData} with the experimental data obtained in 2010 as summarized in Table III of Ref.~\cite{Cheng:SAT}, it is clear that only a few new measurements were available since 2010.  Many existing measurements need further improvement; for example, the uncertainties of $\B(D^0\to f_2K_S\to \pi^+\pi^-K_S)$ and $\B(D^0\to a_2^-\pi^+\to K_SK^-\pi^+)$ are larger or comparable to their central values.  Moreover, as will be discussed in Sec.~\ref{sec:flavorapp}, the existing data of $D^+\to f_2\pi^+, \ov K_2^{*0}\pi^+$ and $K_2^{*0}\pi^+$ are not self consistent.  In other words, the quality of the data needs to be substantially improved.

\begin{table}[t!]
\caption{Experimental branching fractions of various $D\to TP$ decays.  For
simplicity and convenience, we have dropped the mass
identification for $f_2(1270)$, $a_2(1320)$ and $K^*_2(1430)$.  Data are taken from Ref.~\cite{PDG} unless specified otherwise.}
\label{tab:TPData}
 \medskip
\footnotesize{
\begin{ruledtabular}
\begin{tabular}{ l l l }
$\B(D\to TP; T\to P_1P_2)$  & $\B(D\to TP)_{\rm NWA}$ \\
 \hline
 $\B(D^+\to f_2\pi^+; f_2\to\pi^+\pi^-)=(5.0\pm 0.9)\times 10^{-4}$ &
 $\B(D^+\to f_2\pi^+)=({ 8.9\pm 1.6})\times 10^{-4}$ \\
 $\B(D^+\to \ov K^{*0}_2\pi^+; \ov K_2^{*0}\to K^-\pi^+)=(2.3\pm 0.7)\times 10^{-4}$ &
 $\B(D^+\to \ov K^{*0}_2\pi^+)=(6.9\pm 2.1)\times 10^{-4}$ \\
 $\B(D^+\to K^{*0}_2 \pi^+; K_2^{*0}\to K^+\pi^-)=(3.9\pm2.7)\times 10^{-5}$ &  $\B(D^+\to K^{*0}_2 \pi^+)=(1.17\pm0.81)\times 10^{-4}$   \\
 $\B(D^+\to \ov K^{*0}_2 K^+; \ov K_2^{*0}\to K^-\pi^+)=(1.6^{+1.2}_{-0.8})\times 10^{-4}$ & prohibited on shell \\
 \hline
 $\B(D^0\to f_2\pi^0; f_2\to\pi^+\pi^-)=(1.96\pm 0.21)\times 10^{-4}$ &
 $\B(D^0\to f_2\pi^0)=(3.5\pm 0.4)\times 10^{-4}$ \\
 $\B(D^0\to f_2K_S; f_2\to\pi^+\pi^-)=(9^{+10}_{-~6})\times 10^{-5}$ &
 $\B(D^0\to f_2\ov K^0)=(3.2^{+3.6}_{-2.1})\times 10^{-4}$ \\
 $\B(D^0\to f_2K_S; f_2\to\pi^0\pi^0)=(2.3\pm1.1)\times 10^{-4}$ &
 $\B(D^0\to f_2\ov K^0)=(1.6\pm 0.8)\times 10^{-3}$ \\
 $\B(D^0\to K^{*-}_2\pi^+; K_2^{*-}\to K_S^0\pi^-)=(3.4^{+1.9}_{-1.0})\times 10^{-4}$ &
 $\B(D^0\to K^{*-}_2\pi^+)=(2.0^{+1.1}_{ -0.6})\times 10^{-3}$ \\
 $\B(D^0\to K^{*+}_2\pi^-; K_2^{*+}\to K_S^0\pi^+)<3.4\times 10^{-5}$ &
 $\B(D^0\to K^{*+}_2\pi^-)<2.0\times 10^{-4}$ \\
 $\B(D^0\to a_2^-\pi^+; a_2^-\to K_SK^-)=(5\pm5)\times 10^{-6}$ &   $\B(D^0\to a_2^-\pi^+)=(2.0\pm3.9)\times 10^{-4}$ \\
 $\B(D^0\to a_2^+K^-; a_2^+\to K^+K_S)<1.04\times 10^{-4}$ \footnotemark[1]  &  $\B(D^0\to a_2^+K^-)<4.2\times 10^{-3}$ \\
 $\B(D^0\to a_2^-K^+; a_2^-\to K^-K_S)<0.72\times 10^{-4}$ \footnotemark[1] & $\B(D^0\to a_2^-K^+)<2.9\times 10^{-3}$ \\
 \hline
 $\B(D_s^+\to f_2\pi^+; f_2\to\pi^+\pi^-)=(1.09\pm 0.20)\times 10^{-3}$  &
 $\B(D_s^+\to f_2\pi^+)=(1.94\pm 0.36)\times 10^{-3}$ \\
 $\B(D_s^+\to f_2\pi^+; f_2\to\pi^0\pi^0)=(0.80\pm 0.42)\times 10^{-3}$ \footnotemark[2] &
 $\B(D_s^+\to f_2\pi^+)=(2.85\pm 1.50)\times 10^{-3}$ \\
\end{tabular}
\footnotetext[1]{BESIII data taken from Ref.~\cite{BESIII:D0KKKS}.}
\footnotetext[2]{BESIII data taken from Ref.~\cite{BESIII:Dspi+pi0pi0}.}
\end{ruledtabular}}
\end{table}

\section{Physical properties of tensor mesons \label{sec:properties}}

The observed $J^P=2^+$ tensor mesons $f_2(1270)$, $f_2'(1525)$, $a_2(1320)$ and $K_2^*(1430)$ form an SU(3) $1\,^3P_2$ nonet.  The $q\bar q$ content for isodoublet and isovector tensor resonances are obvious.  Just as the $\eta$-$\eta'$ mixing in the pseudoscalar case, the isoscalar tensor states $f_2(1270)$ and $f'_2(1525)$ also have a mixing, and their wave functions are defined by
 \be
 f_2(1270) &=&
{1\over\sqrt{2}}(f_2^u+f_2^d)\cos\theta_{f_2} + f_2^s\sin\theta_{f_2} ~, \non \\
 f'_2(1525) &=&
-{1\over\sqrt{2}}(f_2^u+f_2^d)\sin\theta_{f_2} + f_2^s\cos\theta_{f_2} ~,
 \en
with $f_2^q\equiv q\bar q$.  Since $\pi\pi$ is the dominant decay mode of $f_2(1270)$ whereas $f_2'(1525)$ decays predominantly into $K\ov K$ (see Ref.~\cite{PDG}), it is obvious that this mixing angle should be small.  It is found that $\theta_{f_2}=5.6^\circ$ when the quadratic mass formula for the mixing angle is employed~\cite{PDG,Cheng:mixing}.  Therefore, $f_2(1270)$ is primarily a $(u\bar u+d\bar d)/\sqrt{2}$ state, while $f'_2(1525)$ is dominantly $s\bar s$.

The polarization tensor $\vp_{\mu\nu}$ of a $^3P_2$ tensor meson with $J^{PC}=2^{++}$ satisfies the relations
 \be
 \vp_{\mu\nu}=\vp_{\nu\mu} ~, \qquad \vp^{\mu}_{~\mu}=0 ~, \qquad
 p_\mu \vp^{\mu\nu}=p_\nu\vp^{\mu\nu}=0 ~,
 \en
where $p^\mu$ is the momentum of the tensor meson.  Therefore,
 \be
 \la 0|(V-A)_\mu|T(\vp,p)\ra = a\vp_{\mu\nu}p^\nu+b\vp^\nu_{~\nu} p_\mu=0 ~,
 \en
and hence the decay constant of the tensor meson vanishes identically; that is, the tensor meson cannot be produced from the $V-A$ current.  Nevertheless, a tensor meson can be created from these local currents involving covariant derivatives~\cite{Cheng:LCDAofT}
\be \label{eq:fT}
\la T(P,\lambda)|J_{\mu\nu}(0)|0\ra &=& f_Tm_T^2\epsilon^{*}(\lambda)_{\mu\nu}
~, \non \\
    \langle T(P,\lambda) |J^\perp_{\mu\nu\alpha}(0) |0 \rangle
 &=& -i  f_{T}^\perp m_T
 \left( \epsilon_{\mu\alpha}^{*}(\lambda) P_\nu-\epsilon_{\nu\alpha}^{*}(\lambda) P_\mu \right)
 ~,
\en
where $\lambda$ is the helicity of the tensor meson, and
\be
J_{\mu\nu}(0) &=& \frac{1}{2}
\left( \bar q_1(0)\gamma_\mu i\stackrel{\leftrightarrow}{D}_\nu q_2(0)
     + \bar q_1(0)\gamma_\nu i\stackrel{\leftrightarrow}{D}_\mu q_2(0) \right)
     ~,   \nonumber\\
J^\perp_{\mu\nu\alpha}(0) &=& \bar q_1(0) \sigma_{\mu\nu} i\stackrel{\leftrightarrow}{D}_\alpha q_2(0)
~,
\en
where $\stackrel {\leftrightarrow}{D}_\mu=\stackrel {\rightarrow}{D}_\mu-\stackrel {\leftarrow}{D}_\mu$ with $\stackrel{\rightarrow}{D}_\mu=\stackrel{\rightarrow}\partial_\mu +ig_s A^a_\mu \lambda^a/2$ and $\stackrel{\leftarrow}{D}_\mu=\stackrel{\leftarrow}\partial_\mu -ig_s A^a_\mu \lambda^a/2$.  The decay constants $f_T$ and $f_{T}^\perp$ are scale dependent and they have been evaluated using QCD sum rules at the scale $\mu=1$ GeV~\cite{Cheng:LCDAofT}. We list the results of $f_T$ for later convenience (in units of MeV)
\be
f_T(f_2(1270))= 102\pm 6 ~, &\qquad & f_T(f_2'(1525))=126\pm4 ~, \non \\
f_T(a_2(1320))=107\pm 6 ~, &\qquad & f_T(K_2^*(1430))=118\pm5 ~.
\en

The general expression for the $D\to T$ transition has the form~\cite{Cheng:TP,Wang:2010ni}
\footnote{The $D\to T$ transition form factors defined in Refs.~\cite{Wang:2010ni} and~\cite{Cheng:TP} are different by a factor of $i$. We shall use the former as they are consistent with the normalization of $D\to S$ transition given in Ref.~\cite{CCH}.}
\be \label{eq:FFs}
\langle{T}(p, \lambda)|V_\mu|{ D} (p_D)\rangle
 &=&  \frac{2}{m_D + m_{T}} \varepsilon_{\mu\nu\alpha\beta}
 e_{(\lambda)}^{*\nu}
 p_D^\alpha p^{\beta} V^{DT}(q^2)
 ~, \nonumber \\
 \langle T (p,\lambda)|A_\mu|{ D}(p_D)\rangle
 &=& 2i m_{T} \frac{e^{(\lambda)*}\cdot p_D}{q^2} q_\mu
A_0^{DT}(q^2)+
 i(m_D + m_{T})\left[ e^{(\lambda)*}_{\mu}-\frac{e^{(\lambda)*}\cdot p_D}{q^2} q_\mu\right] A_1^{DT}(q^2) \non \\
&& -i {e^{(\lambda)*} \cdot p_D\over m_D+m_T}\left[p_\mu+(p_D)_\mu-{m_D^2-m_T^2\over q^2}q_\mu\right]
A_2^{DT}(q^2)
~,
 \en
where $q_\mu=(p_D-p)_\mu$ and $e^{*\mu}_{(\lambda)}\equiv \epsilon^{*\mu\nu}(\lambda)(p_D)_\nu/m_D$.  Throughout the paper we will adopt the convention $\varepsilon^{0123}=-1$.  In the ISGW model~\cite{ISGW}, the general expression for the $D\to T$ transition is parametrized as
\be \label{eq:FFinISGW}
 \la T(p,\lambda)|(V-A)_\mu|D(p_D)\ra &=&
 h(q^2)\varepsilon_{\mu\nu\rho\sigma}\epsilon^{*\nu\alpha}p_{D\alpha}(p_D+p)^\rho
 q^\sigma-ik(q^2)\epsilon^*_{\mu\nu}p_D^\nu  \non \\
 && -ib_+(q^2)\epsilon^*_{\alpha\beta}p_D^\alpha p_D^\beta (p_D+p)_\mu
 -ib_-(q^2)\epsilon^*_{\alpha\beta}p_D^\alpha p_D^\beta q_\mu
 ~,
 \en
where the form factor $k$ is dimensionless, and the canonical dimension of $h, b_+$ and $b_-$ is $-2$.
The relations between these two different sets of form factors are given by
\be \label{eq:FFrelations}
&& V^{DT}(q^2)=m_D(m_D+m_T)h(q^2)
~, \qquad A_1^{DT}(q^2)={m_D\over m_D+m_T}k(q^2)
~, \\
&& A_2^{DT}(q^2)=-m_D(m_D+m_T)b_+(q^2)
~, \quad A_0^{DT}(q^2)={m_D\over 2m_T}\left[ k^2(q^2)+(m_D^2-m_T^2)b_+(q^2)+q^2 b_-(q^2) \right]
~. \non
\en

The $D\to T$ transition form factors had been previously evaluated in the ISGW model~\cite{ISGW} and its improved version, ISGW2~\cite{ISGW2}, and the CLFQM~\cite{CCH}.  There were two modern investigations: one was based on the light-cone sum rule approach~\cite{Momeni:2019eow} and the other on the covariant light-front quark model denoted by CLFQM$_b$~\cite{Chen:2021ywv}.  The four form factors $k$, $b_+$, $b_-$ and $h$ defined in Eq.~(\ref{eq:FFinISGW}) for the $D\to T$ transition had been studied in CLFQM and shown in Table~VI of Ref.~\cite{Cheng:SAT}.  It was pointed out in Ref.~\cite{Cheng:TP} that among these four form factors, $k(q^2)$ was particularly sensitive to $\beta_T$, a parameter describing the tensor-meson wave function, and that $k(q^2)$ at zero recoil showed a large deviation from the heavy quark symmetry relation.  It is possible that the very complicated analytic expression for $k(q^2)$ given in Eq.~(3.29) of Ref.~\cite{CCH} is not complete.  To overcome this difficulty, it was advocated in Ref.~\cite{CCH} that one might apply the heavy quark symmetry relation to obtain $k(q^2)$ for $P\to T$ transition (see Eq.~(3.40) of Ref.~\cite{CCH})
\be \label{eq:k}
 k(q^2)=\,m_Pm_{T}\left(1+{m_P^2+m_{T}^2-q^2\over
 2m_Pm_{T}}\right)\left[ h(q^2)-{1\over 2}b_+(q^2)+{1\over 2}b_-(q^2)\right]
 ~.
\en
In other words, the CLFQM results are obtained by first calculating the form factors $h(q^2),b_+(q^2)$ and $b_-(q^2)$ using the covariant light-front approach~\cite{CCH} and $k(q^2)$ from the heavy quark symmetry relation Eq.~(\ref{eq:k}) and then converted them into the form-factor set $V(q^2)$ and $A_{0,1,2}(q^2)$.

Very recently, the $P\to T$ transition form factors $V(q^2)$ and $A_{0,1,2}(q^2)$ were directly evaluated in CLFQM$_b$ in which the issues with self-consistency and Lorentz covariance of the covariant light-front approach were carefully examined and resolved~\cite{Chen:2021ywv}.  It is clear from Table~5 of Ref.~\cite{Chen:2021ywv} that $B\to a_2$ and $B\to K_2^*$ transition form factors obtained in CLFQM and CLFQM$_b$ are consistent with each other, especially for $B\to K_2^*$ transition.

Since the relevant form factor is $A_0^{DT}(q^2)$ in the subsequent study of hadronic $D\to TP$ decays, we exhibit in Table~\ref{tab:FFDtoT} the values of $A_0^{DT}(0)$ in various models. The CLFQM results are obtained from the form factors $h(q^2),b_+(q^2)$ and $b_-(q^2)$ from Table~VI of Ref.~\cite{Cheng:SAT} and $k(q^2)$ from the heavy quark symmetry relation Eq.~(\ref{eq:k}) for the $D\to P$ transition.  Finally, we apply Eq.~(\ref{eq:FFrelations}) to get
$A_0^{DT}(0)$.  Unlike the $B\to P$ transition case, the CLFQM$_b$ results are smaller than CLFQM for various $D\to T$ transitions. Presumably, this means that the heavy quark symmetry relation Eq.~(\ref{eq:k}) has some deviation from the realistic value for $k(q^2)$ as the charm meson is not very heavy. At any rate, we shall take CLFQM$_b$ predictions as the representative values for the covariant light-front approach.

The form-factor $q^2$ dependence in the CLFQM, CLFQM$_b$ and LCSR can be found in Refs.~\cite{Cheng:SAT},~\cite{Chen:2021ywv} and~\cite{Momeni:2019eow}, respectively.  Evidently, the form factors obtained from LCSR are much larger than those in all the other models.  For example, the predicted form factor $A_0^{Da_2}(0)$ in LCSR is larger than that in CLFQM, CLFQM$_b$ and ISGW2 by a factor of 2, 3, and 9, respectively.  This will be tested when we come to the study of $D\to TP$ decays in Sec.~\ref{sec:Results and discussion}.

\begin{table}[t]
\caption{Form factors $A_0^{DT}(q^2)$ for $D\to f_2(1270), a_2(1320),K_2^*(1430)$ transitions at $q^2=0$ in the ISGW2 model~\cite{ISGW2}, the covariant light-front quark models: CLFQM,~\cite{CCH} and CLFQM$_b$~\cite{Chen:2021ywv},  and LCSR~\cite{Momeni:2019eow}. The values of $A_0^{DT}(0)$ in the ISGW2 model are readily obtained from Table VI of~\cite{Cheng:SAT}. The CLFQM results are obtained by first calculating the form factors $h(q^2),b_+(q^2)$ and $b_-(q^2)$ using the covariant light-front approach and $k(q^2)$ from the heavy quark symmetry relation Eq.~(\ref{eq:k}) by setting $P=D$ and then converting them into the form-factor set $V(q^2)$ and $A_{0,1,2}(q^2)$.
}
 \medskip
 \label{tab:FFDtoT}
\begin{ruledtabular}
\begin{tabular}{ l c c c c }
~Transition~~~~~
    & ISGW2 ~\cite{ISGW2}~~~
    & CLFQM~\cite{CCH}~~~
    & CLFQM$_b$~\cite{Chen:2021ywv}~~~
    & LCSR~\cite{Momeni:2019eow}~~~
 \\
    \hline
$D\to f_{2}^q$
    & 0.20
    &  1.10
    &   --
    &  1.92
    \\
$D\to K_2^*$
    & 0.27
    &  1.01
    & $0.68^{+0.06}_{-0.08}$
    &  1.43\footnotemark[1]
    \\
$D\to a_2$
    & 0.20
    & 0.94
    & $0.62^{+0.07}_{-0.07}$
    & 1.80
    \\
$D_s^+\to f_{2}^s$
    & 0.75
    & $0.90$
    & $0.72^{+0.07}_{-0.08}$
    & 1.20
    \\
$D_s^+\to K_2^*$
    & 0.84
    & 0.87
    & $0.58^{+0.05}_{-0.08}$
    & --
    \\
\end{tabular}
\footnotetext[1]{The value of 2.98 for $A_0^{DK_2^*}(0)$ given in Table II of~\cite{Momeni:2019eow} is not consistent with that shown in Fig. 4 of the same reference. The correct value should read 1.43~\cite{private}.}
\end{ruledtabular}
\end{table}

\section{Diagrammatic  approach \label{sec:flavorapp}}

It is known that a least model-dependent analysis of heavy meson decays can be carried out in the so-called topological diagram approach.  In this diagrammatic scenario, all two-body nonleptonic weak decays of heavy mesons can be expressed in terms of six distinct quark diagrams~\cite{Chau,CC86,CC87}: $T$, the external $W$-emission tree diagram; $C$, the internal $W$-emission; $E$, the $W$-exchange; $A$, the $W$-annihilation; $H$, the horizontal $W$-loop; and $V$, the vertical $W$-loop.  These diagrams are classified according to the topologies of weak interactions with all strong interaction effects encoded.  The one-gluon exchange approximation of the $H$ graph is the so-called ``penguin diagram.''  Since given the current data it is premature to consider CP asymmetries in these decays, we ignore both $H$ and $V$ diagrams.

\begin{table}[t]
\caption{Topological amplitudes of $D\to TP$ decays.  The experimental branching fractions denoted by $\B_{\rm NWA}$ are taken from Table~\ref{tab:TPData}.}
\label{tab:DtoTPamp}
\footnotesize{
\begin{ruledtabular}
\begin{tabular}{ l c  c}
Decay & Amplitude & $\B_{\rm NWA}$  \\
\hline
 $D^+\to f_2\pi^+$
& $\frac{1}{\sqrt{2}}V_{cd}^*V_{ud}\cos\theta_{f_2} (T + C' + A + A')+ V_{cs}^*V_{us}\sin\theta_{f_2} C$
& $(8.9\pm 1.6)\times 10^{-4}$ \\
 $D^+\to \ov K^{*0}_2\pi^+$
& $V_{cs}^*V_{ud} (T + C')$
& $(6.9\pm 2.1)\times 10^{-4}$ \\
 $D^+\to K^{*0}_2\pi^+$
& $V_{cd}^*V_{us} (C' + A)$
&  $(1.2\pm 0.8)\times 10^{-4}$ \\
 $D^+\to \ov K^{*0}_2K^+$
& $V_{cs}^*V_{us}T+V_{cd}^*V_{ud}A$
&  prohibited on-shell \\
 \hline
 $D^0\to f_2\pi^0$
& $\frac12 V_{cd}^*V_{ud}\cos\theta_{f_2} (C' - C - E' - E)+ \frac{1}{\sqrt{2}} V_{cs}^*V_{us}\sin\theta_{f_2} C'$
& $(3.5\pm 0.4)\times 10^{-4}$ \\
 $D^0\to f_2\ov K^0$
& $V_{cs}^*V_{ud} \left[
\frac{1}{\sqrt{2}}\cos\theta_{f_2} (C + E) + \sin\theta_{f_2} E' \right]$
&  $(4.6\pm2.7)\times 10^{-4}$ \footnotemark[1]\\
 $D^0\to K^{*-}_2\pi^+$
& $V_{cs}^*V_{ud} (T + E')$
& $(2.0^{+1.1}_{ -0.6})\times 10^{-3}$ \\
 $D^0\to K^{*+}_2\pi^-$
& $V_{cd}^*V_{us} (T' + E)$
&  $<2.0\times 10^{-4}$ \\
 $D^0\to a_2^-\pi^+$
& $V_{cd}^*V_{ud} (T + E')$
&  $(2.0\pm3.9)\times 10^{ -4}$ \\
 $D^0\to a_2^+K^-$
& $V_{cs}^*V_{ud} (T' + E)$
&  $<4.2\times 10^{ -3}$ \\
 $D^0\to a_2^-K^+$
& $V_{cd}^*V_{us} (T + E')$
&  $<2.9\times 10^{ -3}$ \\
\hline
 $D_s^+\to f_2\pi^+$
& $V_{cs}^*V_{ud} \left[
\frac{1}{\sqrt{2}}\cos\theta_{f_2} (A + A') + \sin\theta_{f_2} T \right]$
&  $(2.0\pm 0.4)\times 10^{-3}$ \footnotemark[2]\\
\end{tabular}
\footnotetext[1]{Taken from $D^0\to f_2K_S\to \pi^+\pi^-K_S$ and $D^0\to f_2K_S\to \pi^0\pi^0K_S$ of Table~\ref{tab:TPData}.}
\footnotetext[2]{Taken from $D_s^+\to f_2\pi^+\to \pi^+\pi^-\pi^+$ and $D_s^+\to f_2\pi^+\to \pi^0\pi^0\pi^+$  of Table~\ref{tab:TPData}.}
\end{ruledtabular}}
\end{table}

The topological amplitudes for $D\to TP$ decays have been discussed in Refs.~\cite{ChengTP,Cheng:SAT}. Just as $D\to V\!P$ decays, one generally has two sets of distinct diagrams for each topology. For example, there are two external $W$-emission and internal $W$-emission diagrams, depending on whether the emitted particle is an even-party meson or an odd-parity one.  Following the convention in Refs.~\cite{ChengTP,Cheng:SAT}, we shall denote the primed amplitudes $T'$ and $C'$ for the case when the emitted meson is a tensor meson.  For the $W$-exchange and $W$-annihilation diagrams with the final state $q_1\bar q_2$, the prime amplitude denotes that the even-parity meson contains the quark $q_1$.  Although $T'$ and $C'$ are usually set to zero in the na{\"i}ve factorization approach due to the vanishing vector decay constant of the tensor meson induced from the $V-A$ current, they do receive nonfactorizable contributions which will be elucidated in Sec.~\ref{sec:Factorizable and nonfactorizable amplitudes} below.

The topological amplitudes for $D \to TP$ decays are given in Table~\ref{tab:DtoTPamp}.  We have 15 independent unknown parameters for the 8 topological amplitudes $T,C,E,A$ and $T',C',E',A'$.  It is clear from Table~\ref{tab:DtoTPamp} that we have only 8 available data (some of them being redundant) and three upper limits.  This means that at present, we have more theory parameters than observables.  Moreover, the data for $D^+\to TP$ modes appear not self consistent.  According to the CKM matrix elements associated with each decay mode and the expectation of $|T|\gg|C'|$, it is expected that $\B(D^+\to \ov K^{*0}_2\pi^+)> \B(D^+\to f_2\pi^+)\gg \B(D^+\to K^{*0}_2\pi^+)$.  This hierarchy pattern is not respected by the current data.

\section{Factorization Approach \label{sec:facDTP}}

The diagrammatic approach has been applied quite successfully to hadronic decays of charmed mesons into $PP$ and $V\!P$ final states.
When generalized to the decay modes involving a tensor meson in the final state, it appears that the current data are still insufficient for us to fully extract the information of all amplitudes.    Therefore, we take the na{\"i}ve factorization formalism as a complementary approach to estimate the rates of these decay modes.  In this framework, the $W$-exchange and -annihilation types of contributions will be neglected.

\subsection{Factorizable and nonfactorizable amplitudes \label{sec:Factorizable and nonfactorizable amplitudes}}

The factorizable amplitudes for the $D\to TP$ decays involve the quantities
 \begin{eqnarray} \label{eq:XDSP}
  X^{(D T, P)}
 &\equiv& \langle P(q)| (V-A)_\mu|0\rangle \langle T(p)| (V-A)^\mu|D(p_D)\rangle
 ~, \non \\
  X^{(D P, T)}
 &\equiv& \langle T(q)| (V-A)_\mu|0\rangle \langle P(p)| (V-A)^\mu|D(p_D)\rangle
 ~,
 \end{eqnarray}
with the expression
\be \label{eq:XTP}
 X^{(DT, P)}
 = 2f_P {m_{T}\over m_D} A_0^{DT}(m_P^2)\epsilon^{*\mu\nu}(0)(p_D)_\mu (p_D)_\nu
 ~,
\en
while $ X^{(DP, T)}$ vanishes owing to the fact that the tensor meson cannot be produced through the $V-A$ current. Nevertheless, as shown in Ref.~\cite{Cheng:TP}, beyond the factorization approximation, nonfactorizable contributions proportional to the decay constant $f_T$ defined in Eq.~(\ref{eq:fT}) can be produced from vertex and spectator-scattering corrections
\be
\bar X^{({D} P, T)}= \sqrt{6} f_T  {m_T^2\over m_D\, p_c}\,F_1^{DP}(m_T^2)\epsilon^{*\mu\nu}(0)(p_D)_\mu (p_D)_\nu,
\en
with $p_c$ being the c.m. momentum of either $T$ or $P$ in the $D$ rest frame.

\begin{table}[t]
\caption{Numerical values of the flavor operators $a_i^p(M_1M_2)$ for $M_1M_2=TP$ and $PT$ at the scale $\mu=\ov m_c(\ov m_c)=1.3$ GeV. }
\label{tab:aiTP}
\begin{center}
\begin{tabular}{ l c c | l c c} \hline \hline
 $$~~ & ~~$f_2(1270)\pi$~~ & ~~~$\pi f_2(1270)$~~~ & ~~$$  & $f_2(1270) K$ & $K f_2(1270)$ \\
\hline
 $a_1$~~ & $1.391+0.314i$ & $-0.043+0.021i$~~ & ~~$a_1$ & ~~$1.599+0.928i$ & $-0.074+0.013i$  \\
 $a_2$~~ & $-0.760-0.685i$ & $0.098-0.055i$ &  ~~$a_2$ & $-1.209-2.006i$ & ~~$0.166-0.038i$ \\
 \hline\hline
  $$ & ~~$a_2(1320)\pi$~~ & ~~~$\pi a_2(1320)$~~~ & ~~$$  & ~~$a_2(1320) K$ & $K a_2(1320)$ \\
  \hline
 $a_1$ & $1.372+0.350i$ & $-0.065+0.011i$ & ~~$a_1$ & $1.664+1.576i$ & $-0.089+0.008i$  \\
 $a_2$ & $-0.719-0.760i$ & ~~$0.099-0.055i$ &  ~~$a_2$ & $-1.347-3.402i$ & ~~$0.198-0.028i$ \\
 \hline\hline
   $$ & ~~$K_2^*(1430)\pi$~~ & ~~~$\pi K_2^*(1430)$~~~ & ~~$$  & $K_2^*(1430) K$ & $K K_2^*(1430)$ \\
  \hline
 $a_1$ & $1.254+0.403i$ & $-0.044+0.020i$ & ~~$a_1$ & ~~$1.447-1.394i$~~ & ~~$-0.039+0.098i$~~  \\
 $a_2$ & $-0.467-0.877i$ & ~~$0.101-0.054i$ &  ~~$a_2$ & ~~$-0.880+2.991i$~~ & ~~$0.089-0.221i$~~ \\
\hline \hline
\end{tabular}
\end{center}
\end{table}

The amplitudes  $X^{({D} T,P)}$ and $\bar X^{({D} P,T)}$ can be further simplified by working in the $D$ rest frame  and assuming that $T$ ($P$) moves along the $-z$ ($z$) axis~\cite{Cheng:TP}. In this case, $p_D^\mu=(m_D, 0,0,0)$ and $\epsilon^{*\mu\nu}(0)=\sqrt{2/3}\,\epsilon^{*\mu}(0)\epsilon^{*\nu}(0)$ with $\epsilon^{*\mu}(0)=(p_c,0,0,E_T)^\mu/m_T$ and, consequently,
\be \label{eq:Xf2-2}
X^{({D} T,P)}= 2\sqrt{2\over 3}\,f_P{m_D\over m_T}p_c^2A_0^{D T}(m_P^2), \qquad
\bar X^{(D P,T)}= 2f_Tm_D\, p_cF_1^{DP}(m_T^2).
\en
It is interesting to notice that the expression of  $\bar X^{(D P,T)}$ has a similar structure as $X^{(D P,V)}$
\be
X^{(D P,V)}\equiv \langle V| J'_\mu|0\rangle \langle P| J^\mu|D(p_D)\rangle=2 f_V m_V \epsilon\cdot p_D F_1^{DP}(m_V^2)
\to 2f_Vm_D\, p_cF_1^{DP}(m_V^2)
~.
\en

The color-allowed and color-suppressed tree amplitudes $T,T',C$ and $C'$ then have the expressions (in units of $G_F/\sqrt{2}$)
\be \label{eq:TP_T,C}
&& T= 2 a_1(TP)\sqrt{2\over 3}\,f_P{m_D\over m_T}p_c^2A_0^{D T}(m_P^2)
~, \qquad~ C= 2a_2(TP)\sqrt{2\over 3}\,f_P{m_D\over m_T}p_c^2A_0^{D T}(m_P^2)
~, \non \\
&& T'= 2a_1(PT) f_T  m_D\,p_cF_1^{DP}(m_T^2)
~, \qquad\qquad C'= 2a_2(PT) f_T  m_D\,p_cF_1^{DP}(m_T^2)
~,
\en
where the nonfactorizable amplitudes are dictated by the tensor decay constant $f_T$.

The flavor operator coefficients $a_i(M_1M_2)$ in Eq.~(\ref{eq:TP_T,C}) are basically the Wilson coefficients in conjunction with short-distance nonfactorizable corrections such as vertex corrections and hard spectator interactions. In general, they have the expressions~\cite{BBNS,BN}
\be \label{eq:ai}
  a_1(M_1M_2) &=&
 \left(c_1+{c_2\over N_c}\right)N_1(M_2)  + {c_{2}\over N_c}\,{C_F\alpha_s\over
 4\pi}\Big[V_1(M_2)+{4\pi^2\over N_c}H_1(M_1M_2)\Big], \non \\
   a_2(M_1M_2) &=&
 \left(c_2+{c_1\over N_c}\right)N_2(M_2)  + {c_{1}\over N_c}\,{C_F\alpha_s\over
 4\pi}\Big[V_2(M_2)+{4\pi^2\over N_c}H_2(M_1M_2)\Big],
\en
where $c_i$ are the Wilson coefficients, $C_F=(N_c^2-1)/(2N_c)$ with $N_c=3$, $M_2$ is the emitted meson and $M_1$ shares the same spectator quark as the $D$ meson.  The quantities $V_i(M_2)$ account for vertex corrections, $H_i(M_1M_2)$ for hard spectator interactions with a hard gluon exchange between the emitted meson and the spectator quark of the $D$ meson.  The explicit expressions of $V_{1,2}(M)$ and $H_{1,2}(M_1M_2)$ in QCDF for $B\to TP$ decays are given in Ref.~\cite{Cheng:TP}.  The expression of the quantities $N_i(M_2)$, which are relevant to the factorizable amplitudes, reads
\be \label{eq:Ni}
  N_i(P) = 1, \qquad N_i(T) = 0.
\en
It is obvious that $a_{1,2}(PT)$ vanish in the factorization limit and receive nonfactorizable contributions when the strong interactions are turned on.  Here we generalize the work of Ref.~\cite{Cheng:TP} to $D\to TP$ decays to obtain the relevant flavor operators.  The numerical results for the flavor operators $a_i(M_1M_2)$ with $M_1M_2=TP$ and $PT$ are shown in Table~\ref{tab:aiTP}.


\subsection{Two-body and three-body decays \label{sec:Two-body and three-body decays}}

The decay rate of the $D\to TP$ decay is given by
\be \label{eq:GammaDtoPT}
\Gamma(D\to TP)={p_c\over 8\pi m_D^2}|A(D\to TP)|^2
~,
\en
where the decay amplitude can be read from Table~\ref{tab:DtoTPamp} directly. The tree amplitudes $T, T', C$ and $C'$ are given in Eq.~(\ref{eq:TP_T,C}). Sometimes we write
$A(D\to TP)={\cal M}(D\to TP)\epsilon^{*\mu\nu}(0)(p_D)_\mu (p_D)_\nu$. Then the decay rate is recast to
\be
\Gamma(D\to TP)={p_c^5\over 12\pi m_T^2}\,{m_D^2\over m_T^2}|{\cal M}(D\to TP)|^2
~.
\en
We consider two different $D\to T$ transition form-factor models: the CLFQM$_b$~\cite{Chen:2021ywv} and the  LCSR~\cite{Momeni:2019eow}.

The calculation of three-body decays mediated by tensor resonances is more complicated because of the angular distribution of a tensor meson decaying into two pseudoscalar mesons.
We shall take the decay $D^+\to \ov K_2^{*0}\pi^+\to K^-\pi^+\pi^+$ as an example to illustrate the calculation for the 3-body rate.
Writing $\A_{K_2^*}\equiv A(D^+\to \ov K_2^{*0}\pi^+\to  \pi^+(p_1)K^-(p_2)\pi^+(p_3))$ and following Eq.~(4.16) of Ref.~\cite{Cheng:2020iwk}, we have
\be \label{eq:AK2pi}
\A_{K_2^*} =g^{\ov K_2^{*0}\to K^-\pi^+}\,T_{K_2^*}^{\rm BW}(s_{12}){q^2\over\sqrt{6}}(1-3\cos^2\theta_{13}) A(D^+\to \ov K_2^{*0}(m_{12})\pi^+) +(1\leftrightarrow 3)
~,
\en
where the angular distribution of a tensor meson decaying into two spin-zero particles is governed by  $(1-3\cos^2\theta_{13})$.
In general, the angular momentum distribution is described by the Legendre polynomial $P_J(\cos\theta)$.
In the above equation, $m_{12}=\sqrt{s_{12}}$ is the invariant mass of $K_2^*$,
$\theta_{13}$ is the angle between $\vec{p}_1$ and $\vec{p}_3$ measured in the rest frame of the  resonance $K_2^*$, $q$ is the c.m. momentum given by
\be
q = |\vec{p}_1|=|\vec{p}_2|&=&{\sqrt{[s_{12}-(m_1+m_2)^2][s_{12}-(m_1-m_2)^2]}\over 2m_{12} }
~,
\en
and $T_{K_2^*}^{\rm BW}$ is the relativistic Breit-Wigner line shape for describing the distribution of $K_2^*(1430)$:
\be
T_{K_2^*}^{\rm BW}(s)={1\over s-m^2_{K_2^*}+im_{K_2^*}\Gamma_{K_2^*}(s)}
~,
\en
with the energy-dependent decay width
\be \label{eq:widthK2}
\Gamma_{K_2^*}(s)=\Gamma_{K_2^*}^0\left( {q\over q_0}\right)^5
{m_{K_2^*}\over \sqrt{s}} {X_2^2(q)\over X_2^2(q_0)}
~,
\en
and
\be \label{eq:XJ}
X_2(z)=\sqrt{1\over (z\,r_{\rm BW})^4+3(z\,r_{\rm BW})^2+9}
~,
\en
with $r_{{\rm BW}}\approx 4.0\,{\rm GeV}^{-1}$. In Eq.~(\ref{eq:widthK2}), $q_0$ is the value of $q$ when $m_{12}$ is equal to the $K_2^*$ mass and $\Gamma_{K_2^*}^0$ is the normal width of $K_2^*$.

The explicit expression of $\cos\theta_{13}$ in Eq.~(\ref{eq:AK2pi}) is given by
\be
\cos\theta_{13}=-{1\over 4}\,{Z_1(s_{12},s_{23})\over
|{\vec p}_1||\vec{p}_3|}
~,
\en
with the Zemach form~\cite{Asner:2003gh}
\be
Z_1(s_{12},s_{23})=s_{23}-s_{13}+{(m_D^2-m_3^2)(m_1^2-m_2^2)\over s_{12}}
~.
\en

When $\ov K_2^{*0}$ is on-shell, the decay amplitude of $D^+\to \ov K_2^{*0}\pi^+$ is given by $V_{cs}^*V_{ud}(T+C')$ (see Table~\ref{tab:DtoTPamp}). Hence,
\be \label{eq:Af2pi_bar}
A(D^+\to \ov K_2^{*0}(m_{12})\pi^+) &=&
  \frac{G_F}{\sqrt{2}}V_{cs}^*V_{ud}\Big[ 2a_1(K_2^*\pi)\sqrt{2\over 3}f_\pi{m_D\over m_{12}} \tilde p_c^2 A_0^{DK_2^*}(m_\pi^2)  \non \\
  &&~~+2a_2(\pi K_2^*)f_{K_2^*}m_D\,\tilde p_c  F_1^{D\pi}(s_{12}) \Big]
  ~,
\en
where use of Eq.~(\ref{eq:TP_T,C}) has been made and $\tilde p_c$ is the c.m. momentum of $\pi^+$ or $\ov K_2^{*0}$ with the invariant mass $m_{12}$ in the rest frame of $D^+$
\be
\tilde p_c={1\over 2m_D}\sqrt{[m_D^2-(m_{12}+m_3)^2][m_D^2-(m_{12}-m_3)^2]}
~.
\en
Note that in the rest frame of $K_2^*$, the momentum of $\pi^+(p_3)$ reads
\be
|\vec{p}_3| = \left({(m_D^2-m_{12}^2-m_3^2)^2\over 4m_{12}^2} - m_3^2\right)^{1/2}
~.
\label{eq: p3}
\en
We see that $\tilde p_c$ is related to $|\vec p_3|$  through the relation $\tilde p_c=(m_{12}/m_D)|\vec{p}_3|$.

Finally, the 3-body decay rate reads
\be \label{eq:Gamma_K2st}
&&\Gamma(D^+\to \ov K_2^{*0}\pi^+ \to \pi^+K^-\pi^+) \non
\\
&&={1\over 2}\,{1\over(2\pi)^3 32 m_D^3}\int_{(m_K+m_\pi)^2}^{(m_D-m_\pi)^2}ds_{12}\int_{(s_{23})_{\rm min}}^{(s_{23})_{\rm max}}ds_{23} \, |\A_{K_2^*}|^2 \non \\
&&={1\over 2}\,{1\over(2\pi)^3 32 m_D^3}\int_{(m_K+m_\pi)^2}^{(m_D-m_\pi)^2}ds_{12}\int_{(s_{23})_{\rm min}}^{(s_{23})_{\rm max}}ds_{23}\Bigg\{  {\left|g^{\ov K_2^*\to K^-\pi^+}\right|^2 F(s_{12},m_{K_2^*})^2\over (s_{12}-m^2_{K_2^*})^2+m_{K_2^*}^2\Gamma_{K_2^*}^2(s_{12})}   \\
&&~~\times {q^4\over 6}\left(1-{3\over 16}{Z_1^2(s_{12},s_{23})\over q^2\tilde p_c^2}\,{s_{12}\over m_D^2} \right)^2 \left|A(D^+\to \ov K_2^{*0}(m_{12})\pi^+)\right|^2 +(s_{12}\leftrightarrow s_{23}) + {\rm interference} \Bigg\}
~, \non
\en
where the factor of 1/2 accounts for the identical-particle effect.
The coupling constant $g^{\ov K_2^*\to K^-\pi^+}$ is determined from the measured width of
$\ov K_2^*(1430)\to K^-\pi^+$ through the relation
\be \label{eq:K2toKpi}
\Gamma_{\ov K_2^*(1430)\to K^-\pi^+}={q_0^5\over 60\pi m_{K_2^*}^2}\left|g^{\ov K_2^{*0}\to K^-\pi^+}\right|^2
~.
\en
When $K_2^*$ is off the mass shell, especially when $s_{12}$ is approaching the upper bound of $(m_D-m_\pi)^2$, it is necessary to account for the off-shell effect. For this purpose, we have followed Ref.~\cite{Cheng:FSI} to introduce a form factor $F(s,m_R)$ parametrized as
\be \label{eq:FF for coupling}
F(s,m_R)=\left( {\Lambda^2+m_R^2 \over \Lambda^2+s}\right)^n
~,
\en
with the cutoff $\Lambda$ not far from the resonance,
\be
\Lambda=m_R+\beta\Lambda_{\rm QCD}
~,
\en
where the parameter $\beta$ is expected to be of order unity. We shall use $n=1$, $\Lambda_{\rm QCD}=250$ MeV and $\beta=1.0\pm0.2$ in subsequent calculations.

\begin{table}[t]
\caption{Topological amplitudes and branching fractions of $D\to TP$ decays.  Theory predictions are made with two different form-factor models: (I) the CLFQM$_b$ and (II) the LCSR, where the mixing angle $\theta_{f_2} = 5.6^\circ$ has been used.  For simplicity and convenience, we have dropped the mass identification for $f_2(1270)$, $a_2(1320)$ and $K^*_2(1430)$. Branching fractions denoted by $\B_{\rm NWA}$ are taken from Table~\ref{tab:TPData}.
} \label{tab:DtoTPtheory}
\vskip 0.3cm
\footnotesize{
\begin{ruledtabular}
\begin{tabular}{ l  c c c}
Decay & Model I & Model II & $\B_{\rm NWA}$ \\
\hline
 $D^+\to f_2\pi^+$
& $8.9\times 10^{-5}$
& $6.4\times 10^{-4}$ & $(8.9\pm 1.6)\times 10^{-4}$ \\
 $D^+\to \ov K^{*0}_2\pi^+$
& $1.1\times 10^{-3}$
& $3.0\times 10^{-3}$ & $(6.9\pm 2.1)\times 10^{-4}$ \\
 $D^+\to K^{*0}_2\pi^+$
& $5.6\times 10^{-7}$
& $5.6\times 10^{-7}$ & $(1.2\pm 0.8)\times 10^{-4}$ \\
 $D^+\to a_2^+\ov K^0$
& $2.9\times 10^{-5}$
& $1.3\times 10^{-4}$ & $$ \\
 \hline
 $D^0\to f_2\pi^0$
& $8.3\times 10^{-6}$
& $6.0\times 10^{-5}$ & $(3.5\pm 0.4)\times 10^{-4}$ \\
 $D^0\to f_2\ov K^0$
& $7.8\times 10^{-5}$
& $5.3\times 10^{-4}$ & $(4.6\pm2.7)\times 10^{-4}$ \\
 $D^0\to K^{*-}_2\pi^+$
& $1.8\times 10^{-4}$
& $7.7\times 10^{-4}$ & $(2.0^{+1.1}_{ -0.6})\times 10^{-3}$ \\
 $D^0\to K^{*+}_2\pi^-$
& $3.9\times 10^{-8}$
& $3.9\times 10^{-8}$ & $<2.0\times 10^{-4}$ \\
 $D^0\to a_2^-\pi^+$
& $3.0\times 10^{-5}$
& $2.5\times 10^{-4}$ & $(2.0\pm3.9)\times 10^{ -4}$ \\
 $D^0\to a_2^+K^-$
& $6.1\times 10^{-6}$
& $1.7\times 10^{-6}$ & $<4.2\times 10^{ -3}$ \\
 $D^0\to a_2^-K^+$
& $9.8\times 10^{-8}$
& $2.2\times 10^{-7}$ & $<2.9\times 10^{ -3}$ \\
\hline
 $D_s^+\to f_2\pi^+$
& $2.9\times 10^{-5}$
& $7.7\times 10^{-4}$ & $(2.0\pm 0.4)\times 10^{-3}$ \\
%
%
 $D_s^+\to K_2^{*0}\pi^+$
& $2.3\times 10^{-5}$ & $$  \\
\end{tabular}
\end{ruledtabular}}
\end{table}

\begin{table}[!]
\caption{Same as Table~\ref{tab:DtoTPtheory} except for $D\to TP\to P_1P_2P$ decays.
}
\label{tab:DtoTP3bdoy:theory}
\vskip 0.3cm
\footnotesize{
\begin{ruledtabular}
\begin{tabular}{l  c c c }
$D\to TP; T\to P_1P_2$  & Model I & Model II & Experiment \\ \hline
 $D^+\to f_2\pi^+; f_2\to\pi^+\pi^-$ & $5.7\times 10^{-5}$ & $4.2\times 10^{-4}$ & $(5.0\pm 0.9)\times 10^{-4}$
 \\
$D^+\to \ov K_2^{*0}\pi^+; \ov K_2^{*0}\to K^-\pi^+$ & $4.2\times 10^{-4}$ & $1.3\times 10^{-3}$  & $(2.3\pm0.7)\times 10^{-4}$ \\
$D^+\to K_2^{*0}\pi^+; K_2^{*0}\to K^+\pi^-$ & $1.6\times 10^{-7}$ & $1.6\times 10^{-7}$ & $(3.9\pm2.7)\times 10^{-5}$ \\
$D^+\to \ov K_2^{*0}K^+; \ov K_2^{*0}\to K^-\pi^+$ & $1.4\times 10^{-6}$ & $8.2\times 10^{-6}$ & $(1.6^{+1.2}_{-0.8})\times 10^{-4}$
\\
$D^+\to a_2^{+} \ov K^0;  a_2^{+}\to K^+\ov K^0$ & $3.6\times 10^{-5}$ & $2.0\times 10^{-4}$  & \\
$D^+\to a_2^{+}\ov K^0;  a_2^{+}\to \eta\pi^+$ & $1.4\times 10^{-4}$ & $7.7\times 10^{-4}$  & \\
\hline
 $D^0\to f_2\pi^0; f_2\to\pi^+\pi^-$ & $4.3\times 10^{-6}$ & $3.2\times 10^{-5}$  & $(1.96\pm 0.21)\times 10^{-4}$ \\
 $D^0\to f_2\ov K^0; f_2\to\pi^+\pi^-$ & $1.5\times 10^{-4}$ & $1.0\times 10^{-3}$  & $(1.8^{+2.0}_{-1.2})\times 10^{-4}$  \\
 $D^0\to f_2\ov K^0; f_2\to\pi^0\pi^0$ & $7.6\times 10^{-5}$ & $5.1\times 10^{-4}$  & $(4.6\pm2.2)\times 10^{-4}$\\
 $D^0\to K^{*-}_2\pi^+; K_2^{*-}\to \ov K^0\pi^-$ & $6.3\times 10^{-5}$ & $2.7\times 10^{-4}$  & $(6.8^{+3.8}_{-2.0})\times 10^{-4}$ \\
 $D^0\to K^{*+}_2\pi^-; K_2^{*+}\to K^0\pi^+$ & $1.2\times 10^{-8}$ & $1.2\times 10^{-8}$  & $<6.8\times 10^{-5}$ \\
 $D^0\to a_2^-\pi^+; a_2^-\to K^0K^-$ & $1.2\times 10^{-6}$ & $9.5\times 10^{-6}$  & $(1\pm1)\times 10^{-5}$  \\
 $D^0\to a_2^+K^-; a_2^+\to K^+\ov K^0$ & $2.7\times 10^{-7}$ & $2.7\times 10^{-7}$  & $<2.08\times 10^{-4}$  \\
 $D^0\to a_2^-K^+; a_2^-\to K^-K^0$ & $9.8\times 10^{-9}$ & $5.8\times 10^{-8}$  & $<1.44\times 10^{-4}$ \\
 \hline
 $D_s^+\to f_2\pi^+; f_2\to\pi^+\pi^-$ & $1.7\times 10^{-5}$ & $4.7\times 10^{-5}$  & $(1.09\pm 0.20)\times 10^{-3}$  \\
 $D_s^+\to f_2\pi^+; f_2\to\pi^0\pi^0$ & $7.2\times 10^{-6}$ & $2.0\times 10^{-5}$ & $(0.80\pm 0.42)\times 10^{-3}$ \\
\end{tabular}
\end{ruledtabular}
}
\end{table}

\subsection{Results and discussion \label{sec:Results and discussion}}

Branching fractions of two-body $D\to TP$ and three-body $D\to TP\to P_1P_2P$ decays are displayed in Tables~\ref{tab:DtoTPtheory} and \ref{tab:DtoTP3bdoy:theory}, respectively, using the factorization approach with $W$-exchange and $W$-annihilation being neglected.  Theory predictions are made with two different form-factor models: the CLFQM$_b$ and the LCSR.

It appears from Tables~\ref{tab:DtoTPtheory} and \ref{tab:DtoTP3bdoy:theory} that form factors $A_0^{DT}$ based on LCSR give a better agreement with experiment, keeping in mind that we have neglected contributions from both $W$-exchange and $W$-annihilation.  Since the decays $D^+\to \ov K_2^{*0}\pi^+$ and $D^+\to a_2^+\ov K^0$ do not receive $W$-annihilation contributions, they are ideal for testing the factorization hypothesis. Our prediction of $\B(D^+\to \ov K_2^{*0}\pi^+)=(1.1-3.0)\times 10^{-3}$ seems to be too large compared to the experimental value of $(6.9\pm2.1)\times 10^{-4}$.  As discussed in Sec. IV, the data for $D^+\to TP$ modes do not appear to be self consistent.  Since the mixing angle $\theta_{f_2}$ is small, to a good approximation with negligible $W$-annihilation compared to external $W$-emission, it is expected that $\B(D^+\to f_2\pi^+)/\B(D^+\to \ov K_2^{*0}\pi^+)\approx (\sin\theta_C)^2/2=0.025$\,.  However,
the current data indicate the other way around, $\B(D^+\to f_2\pi^+)\gsim B(D^+\to \ov K_2^{*0}\pi^+)$.  The mode $D^+\to K_2^{*0}\pi^+$ is doubly Cabibbo-suppressed and hence one would expect $\B(D^+\to \ov K_2^{*0}\pi^+)\gg \B(D^+\to K_2^{*0}\pi^+)$, which is not borne out by current experiments.  We hope that the quality of data in the $D^+$ sector will be improved in the near future.

For $D^0\to TP$ decays, LCSR form factors lead to better agreement between theory and experiment. Nevertheless, the role of $W$-exchange and $W$-annihilation should be investigated. Especially, external $W$-emission in $D_s^+\to f_2\pi^+$ is suppressed by the small mixing angle $\theta_{f_2}$, but $W$-annihilation is not.

As noticed in passing, the decay $D^+\to\ov K_2^{*0}K^+$ is prohibited for physical $K_2^*$ and $K$ states. Nevertheless, the 3-body decay $D^+\to\ov K_2^{*0}K^+\to K^-\pi^+K^+$
can proceed owing to the finite width of $K_2^*$. Our calculation shows that the predicted rate is too small by one to two orders of magnitude compared to experiment (see Table~\ref{tab:DtoTP3bdoy:theory}).

An inspection of Tables~\ref{tab:DtoTPtheory} and \ref{tab:DtoTP3bdoy:theory} may lead the reader to wonder why the calculations in these two tables seem not to respect the NWA given by
Eq.~(\ref{eq:fact}). For example, $\B(D^+\to \ov K_2^{*0}\pi^+)=3.0\times 10^{-3}$ and
$\B(D^+\to \ov K_2^{*0}\pi^+\to K^-\pi^+\pi^+)=1.3\times 10^{-3}$ obtained using the LCSR form factor do not satisfy the factorization relation
\be
\B(D^+\to \ov K_2^{*0}\pi^+\to K^-\pi^+\pi^+)=\B(D^+\to \ov K_2^{*0}\pi^+)\B(\ov K_2^{*0}\to K^-\pi^+)
~,
\en
with $\B(\ov K_2^{*0}\to K^-\pi^+)={2\over 3}(0.499\pm 0.0012)$. This is ascribed to the finite-width effect which we are going to discuss in the next section.

\section{Finite Width Effects \label{sec:finitewidth}}

The finite-width effect is accounted for by the quantity $\eta_R$ defined by~\cite{Cheng:2020mna,Cheng:2020iwk}
\be \label{eq:eta}
\eta_{_R}\equiv \frac{\Gamma(D\to RP_3\to P_1P_2P_3)_{\Gamma_R\to 0}}{\Gamma(D\to RP_3\to P_1P_2P_3)}=\frac{\Gamma(D\to RP_3)\B(R\to P_1P_2)}{\Gamma(D\to RP_3\to P_1P_2P_3)}=1+\delta
~,
\en
so that the deviation of $\eta_{_R}$ from unity measures the degree of departure from the NWA when the resonance width is finite.  It is na{\"i}vely expected that the correction $\delta$ will be of order $\Gamma_R/m_R$. After taking into account the finite-width effect $\eta_R$ from the resonance, the branching fraction of the quasi-two-body decay reads~\cite{Cheng:2020iwk},
\be \label{eq:BRofDtoTP}
\B(D\to RP)=\eta_R\B(D\to RP)_{\rm NWA}=\eta_R{\B(D\to RP_3\to P_1P_2P_3)_{\rm expt}\over
\B(R\to P_1P_2)_{\rm expt}}
~,
\en
where $\B(D\to R P)_{\rm NWA}$ denotes the branching fraction obtained from Eq.~(\ref{eq:fact}) valid in the NWA.

\begin{table}[t]
\caption{A summary of the $\eta_R$ parameter for  tensor resonances produced in the three-body $D$ decays calculated using (I) the CLFQM$_b$ and (II) the LCSR for $D\to T$ transition form factors. The masses and widths of tensor resonances are taken from Ref.~\cite{PDG}.
}
\label{tab:eta}
\vskip 0.15cm
\label{tab:eta}
\footnotesize{
\begin{ruledtabular}
\begin{tabular}{ l l c c c c c}
 Resonance~~~ & ~$D\to Rh_3\to h_1h_2h_3$ ~~~ & ~$\Gamma_R$ (MeV)~~ & ~$m_R$ (MeV)~ &  $\Gamma_R/m_R$ & $\eta_R$ (I) & $\eta_R$ (II)\\
\hline
$K_2^*(1430)$ & $D^+\to \ov K_2^{*0}\pi^+\to K^-\pi^+\pi^+$ & ~$109\pm5$~~ & ~$1432.4\pm1.3$~~ & $0.076\pm0.002$ & 0.835 & 0.787 \\
$a_2(1320)$ & $D^+\to a_2^{+} \ov K^0\to K^+\ov K^0\ov K^0 $ & ~$107\pm5$~~ & ~$1318.2\pm0.6$~~ & $0.081\pm0.004$ & 0.388 & 0.370 \\
 & $D^+\to a_2^{+}\ov K^0 \to \eta\pi^+\ov K^0 $ & ~$107\pm5$~~ & ~$1318.2\pm0.6$~~ & $0.081\pm0.004$ & 0.300 & 0.280 \\
\end{tabular}
\end{ruledtabular} }
\end{table}

We calculate the $\eta_R$ parameters for  tensor resonances produced in the three-body $D$ decays using (I) the CLFQM$_b$ and (II) the LCSR for $D\to T$ transition form factors.  The results are displayed in Table~\ref{tab:eta}. We only consider the $D^+$ decays as the three-body modes listed in Table~\ref{tab:eta} are not contaminated by the $W$-annihilation amplitude and hence the calculated finite width effects are more trustworthy.  The $\eta_R$ parameters for various resonances produced in the three-body $B$ decays have been evaluated in
Refs.~\cite{Cheng:2020mna,Cheng:2020iwk}.

We have checked analytically and numerically that $\eta_R\to 1$ in the narrow width limit as it should be. To see this, we consider the 3-body decay $D^+\to \ov K_2^{*0}\pi^+\to K^-\pi^+\pi^+$ mentioned in Sec.~\ref{sec:Two-body and three-body decays} as an illustration.
The angular distribution in Eq.~(\ref{eq:Gamma_K2st}) has the expression (see Eq.~(4.25) of Ref.~\cite{Cheng:2020iwk})
\be
\int_{(s_{23})_{\rm min}}^{(s_{23})_{\rm max}}ds_{23}(1-3\cos^2\theta_{13})^2={16\over 5}{m_D\over m_{12}}q\,\tilde p_c.
\en
In the narrow width limit of $\Gamma_{K^*_2}$, we have
\be
{m_{K_2^*}\Gamma_{K_2^*}(s)\over (s-m^2_{K_2^*})^2+m_{K_2^*}^2\Gamma_{K_2^*}^2(s)} \xlongrightarrow[]{\; \Gamma_{K_2^*}\to 0 \;}\pi\delta(s-m_{K_2^*}^2).
\en
Under the NWA, $|g^{\ov K_2^{*0}\to K^-\pi^+}|^2/\Gamma_{K_2^{*0}}$ is finite as it is proportional to the branching fraction $\B(\ov K_2^{*0}\to K^-\pi^+)$. Due to the Dirac $\delta$-function in the above equation, we have $s_{12}\to m_{K_2^*}^2$ in the zero width limit. As a result, $\tilde p_c\to p_c$ and $q\to q_0$. Likewise, the second term in Eq.~(\ref{eq:Gamma_K2st}) with the replacement $s_{12}\leftrightarrow s_{23}$ has a similar expression. However, the interference term vanishes in the NWA due to different $\delta$-functions.
From Eqs. (\ref{eq:K2toKpi}), (\ref{eq:GammaDtoPT}) and (\ref{eq:Gamma_K2st}),
we are led to the desired factorization relation
\be \label{eq:factorization_f2}
\Gamma(D^+\to \ov K_2^{*0}\pi^+\to K^-\pi^+\pi^+) \xlongrightarrow[]{\; \Gamma_{K_2^*}\to 0 \;}
 \Gamma(D^+\to \ov K_2^{*0}\pi^+) \B(\ov K_2^{*0}\to K^-\pi^+),
\en
in the zero width limit.

It is evident from Table~\ref{tab:eta} that the finite-width effect is quite significant in the tensor meson production in $D$ decays, $\eta_{K_2^*}\sim 0.79-0.84$ and $\eta_{a_2}\sim 0.37-0.39$ for $a_2\to K\ov K$ and of order $0.28-0.30$ for $a_2\to \eta\pi$.  Recall that in three-body $B$ decays, $\eta_R$ is found to be $0.972\pm0.001$ and $1.004^{+0.001}_{-0.002}$\,, respectively, for $f_2$ and $K_2^*$~\cite{Cheng:2020iwk}.  This means that the deviation of $\B(D\to T\!P)_{\rm NWA}$ from the realistic branching fraction of quasi-two-body decay $\B(D\to T\!P)$ is large and needs to be corrected through Eq.~(\ref{eq:BRofDtoTP}).  If the mass of the $D$ meson were adjusted slightly higher by a few hundred MeV, the value of $\eta_R$ would quickly go up to values around unity.  The same effect would be witnessed if one scaled down the mass and width of the resonance particle.

\section{Conclusions \label{sec:conclusions}}

In this work we have examined the quasi-two-body $D\to TP$ decays and the three-body $D$ decays proceeding through intermediate tensor resonances. Our main results are:

\begin{itemize}

\item
Two new model calculations of $D\to T$ transition form factors are available very recently:
one is based on QCD sum rules (LCSR) and the other on the covariant light-front quark model (CLFQM). Our calculations of two-body and three-body $D$ decays in the factorization approach are based on these two form-factor models.
It appears that form factors based on LCSR give a better agreement with current data.

\item
Denoting the primed amplitudes $T'$ and $C'$ for the case when the emitted meson is a tensor meson, it is na{\"i}vely expected that $T'=C'=0$  as the tensor meson cannot be produced through the $V-A$ current.
Nevertheless, beyond the factorization approximation, contributions proportional to the tensor decay constant $f_T$ can be produced from vertex and hard
spectator-scattering corrections.

\item
We have studied the flavor operator coefficients $a_{1,2}(M_1M_2)$ for $M_1M_2=TP$ and $PT$ within the framework of QCD factorization.  It follows that $a_i(PT)$ and $a_i(TP)$ are very different as the former does not receive factorizable contributions.

\item
We have studied the finite-width effects due to tensor mesons and found it is quite significant in the tensor meson production in $D$ decays, $\eta_{K_2^*}\sim 0.79-0.84$ and $\eta_{a_2}\sim 0.37-0.39$ for $a_2\to K\ov K$ and of order $0.28-0.30$ for $a_2\to \eta\pi$, while in three-body $B$ decays, $\eta_R$ is found to be close to unity for $f_2$ and $K_2^*$.
This implies that the deviation of $\B(D\to T\!P)_{\rm NWA}$, obtained from narrow width approximation, from the realistic branching fraction of quasi-two-body decay $\B(D\to  T\!P)$ is large and needs to be corrected by taking $\eta_R$ into account.

\item
Although the decay $D^+\to\ov K_2^{*0}K^+$ is prohibited for physical $K_2^*$ and $K$ states, the 3-body decay $D^+\to\ov K_2^{*0}K^+\to K^-\pi^+K^+$
can proceed owing to the finite width of $K_2^*$. However, our calculation shows that the predicted rate is too small by one to two orders of magnitude compared to experiment.

\item
We have 15 unknown parameters for the 8 topological amplitudes $T,C,E,A$ and $T',C',E',A'$. However, there are only 8 available data (some of them being redundant) to fit and three upper limits.  At present, we are not able to extract topological amplitudes as
we have more theory parameters than observables.

\item
The current data of $D^+\to f_2\pi^+$ and $D^+\to \ov K_2^{*0}\pi^+$  are not self consistent and need to be clarified in the future. In general, the quality of data for $D\to TP\to P_1P_2P$ needs to be substantially improved.

\end{itemize}

\section*{Acknowledgments}

This research was supported in part by the Ministry of Science and Technology of R.O.C. under Grant Nos.~MOST-107-2119-M-001-034, MOST-110-2112-M-001-025 and MOST-108-2112-M-002-005-MY3, the National Natural Science Foundation of China under Grant No. 11347030, the Program of Science and Technology Innovation Talents in Universities of Henan Province 14HASTIT037.

\vskip 2.5cm


\begin{thebibliography}{99}
\bibitem{private}
We wish to thank Reza Khosravi for this clarification.
\end{thebibliography}
\end{document}